\documentclass[twocolumn,preprintnumbers,amsmath,amssymb]{revtex4}

\usepackage{graphicx}
\usepackage{dcolumn}
\usepackage{bm}

\begin{document}


\title{Vesignieite BaCu$_3$V$_2$O$_8$(OH)$_2$ as a Candidate Spin-1/2 Kagome 
Antiferromagnet}

%

\author{Yoshihiko Okamoto, Hiroyuki Yoshida and Zenji Hiroi}
\affiliation{
Institute for Solid State Physics, University of Tokyo, Kashiwa, Chiba 277-8581, Japan \\
}

\date{\today}

\begin{abstract}
A polycrystalline sample of vesignieite BaCu$_3$V$_2$O$_8$(OH)$_2$ comprising a nearly 
ideal kagome lattice composed of Cu$^{2+}$ ions carrying spin 1/2 has been synthesized 
and studied by magnetization and heat capacity measurements. 
Magnetic susceptibility shows a neither long range order, a spin glass transition nor a spin gap down to 2 K, in spite of a moderately strong antiferromagnetic interaction of $J$/$k_{\textrm{B}}$ = 53 K between nearest-neighbor spins. A broad peak observed at a temperature corresponding
to 0.4$J$ in intrinsic magnetic susceptibility indicates 
a marked development of the short-range order. 
The ground state of vesignieite is probably
a gapless spin liquid or is accompanied by a very small gap less than $\sim$ $J$/30.

\end{abstract}

\maketitle

Spin-1/2 kagome antiferromagnets (KAFMs) with strong quantum fluctuations are one of 
the most intriguing playgrounds in the field of highly frustrated magnetism. 
In spite of numerous theoretical and experimental studies in the last decade, 
the nature of the ground state of the spin-1/2 KAFM 
remains an open question. 
One of the most possible ground state proposed by exact diagonalization studies 
is a nonmagnetic spin-liquid state with a small spin gap estimated to be on the 
order of $J$/20, which is filled with a continuum of singlet excitations, 
where $J$ is the nearest-neighbor exchange coupling constant~\cite{1}. 
On the other hand, we need experiments on an actual material to realize the spin-1/2 KAFM.
However, only a few model systems are known at present, where
some extrinsic factors like three-dimensionality, disorder effects,
and lattice distortion have
prevented us from deducing the intrinsic properties of the KAFM 
at low temperatures.

Two Cu$^{2+}$ minerals, namely, herbertsmithite ZnCu$_3$(OH)$_6$Cl$_2$~\cite{4} and 
volborthite Cu$_3$V$_2$O$_7$(OH)$_2$$\cdot$2H$_2$O~\cite{3}, are known as
model systems for the spin-1/2 KAFM. 
They have slightly different kagome lattices of Cu$^{2+}$ ions.
Herbertsmithite crystallizes in a rhombohedral structure with the 
space group $R$\={3}$m$, comprising an ideal arrangement of 
Cu$^{2+}$ ions in the kagome geometry, so as to be named a ``structurally 
perfect kagome compound''~\cite{4}. 
Intensive experiments have established the absence of 
long-range order or spin glass transition down to 50 mK, 
in spite of the presence of strong antiferromagnetic interactions 
between Cu spins, as evidenced by the large Curie-Weiss 
constant $\theta_{\textrm{W}}$ = $-$300 K~\cite{5}. 
However, a neutron diffraction study revealed that 
$\sim$ 10\% of Cu$^{2+}$ sites in the kagome plane are occupied 
by Zn$^{2+}$ ions~\cite{6}. 
This means that 30\% of Zn sites bridging the kagome layers are occupied 
by Cu$^{2+}$ ions, which cause nearly free impurity spins with weak 
magnetic interactions. 
The existence of these impurity spins has also been supported by 
the observations of a large Shottky-type contribution in 
heat capacity, divergent magnetic susceptibility toward $T$ = 0, and 
magnetization curves with downward curvatures at low temperatures~\cite{7}. 
The presence of these extrinsic contributions has made it difficult to
determine the intrinsic ground state of the kagome lattice in herbertsmithite.

Another model system is volborthite Cu$_3$V$_2$O$_7$(OH)$_2$$\cdot$2H$_2$O,
which crystallizes in a monoclinic structure with the space group $C$2/$m$~\cite{9}.
There are two Cu sites, namely, Cu1 and Cu2, in the kagome plane. 
Thus, the kagome lattice is slightly distorted, consisting of corner-shared isosceles
Cu1-Cu2-Cu2 triangles, where the distances between two Cu atoms 
are 3.03 \AA \ (Cu1-Cu2) and 2.94 \AA \ (Cu2-Cu2)~\cite{9}. 
This structural distortion should give rise to spatial anisotropy
in $J$ in the kagome lattice, resulting in a $J_1$-$J_2$ kagome system 
rather than in a uniform one.
The magnitude of the anisotropy has not yet been determined thus far, but was estimated 
to be less than 20\% by theoretical analysis~\cite{10}. 
The kagome layers in volborthite are well separated by the V$_2$O$_7$ group
and water molecules, resulting in a good two-dimensionality.
Unlike herbertsmithite, volborthite
does not suffer from such antisite disorder between Cu$^{2+}$ and Zn$^{2+}$ ions, 
because V$^{5+}$ ions cannot replace Cu$^{2+}$ ions:
a mutual exchange between Cu$^{2+}$ and V$^{5+}$ ions is unfavorable in
terms of ionic radius and Madelung energy.
In spite of the strong antiferromagnetic interactions revealed by the large negative
$\theta_{\textrm{W}}$ = $-$115 K, volborthite 
shows no long-range magnetic order down to 60 mK, 
but exhibits a small spin glass transition at 1.1 K, which is likely caused 
by impurity spins~\cite{111, 11}. 
Magnetic susceptibility exhibits a broad maximum 
at $T$ $\sim$ 0.25$J$/$k_{\textrm{B}}$, indicative
of the development of a short range order, and approaches a large finite value at 60 mK, 
suggesting a gapless or a very small spin-gapped (less than $J$/1500) 
ground state~\cite{11}. 
Very recently, a polycrystalline sample with much better quality 
has been prepared, and an intriguing phenomenon called ``magnetization steps'' was 
observed in magnetization curves at low temperatures~\cite{11}. 
However, it is still unclear whether these results on volborthite are intrinsic to the
spin-1/2 KAFM or related to deviations from the ideal model.
Therefore, it is necessary to search for alternative, more ideal kagome compounds in nature.
In this letter, we report that vesignieite 
BaCu$_3$V$_2$O$_8$(OH)$_2$ can be a suitable compound for realizing spin-1/2 KAFMs. 
The thermodynamic properties of vesignieite are studied
and compared with those of the previous kagome compounds.

\begin{figure}
\includegraphics[width=7cm]{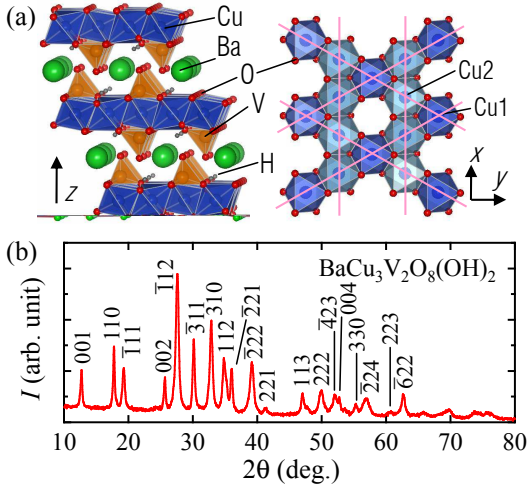}
\caption{(a) Crystal structure of vesignieite BaCu$_3$V$_2$O$_8$(OH)$_2$ viewed 
along the $a$ axis (left) and perpendicular to the $ab$ plane (right). 
(b) Powder XRD pattern of a polycrystalline sample of vesignieite
taken at room temperature. 
Peak indices are given by assuming a monoclinic unit cell of 
$a$ = 10.273 \AA, $b$ = 5.907 \AA, $c$ = 7.721 \AA \ and $\beta$ = 116.29$^{\circ}$.
}
\label{F1}
\end{figure}

Vesignieite BaCu$_3$V$_2$O$_8$(OH)$_2$ is a natural mineral reported 
about a half century ago~\cite{12}. 
It crystallizes in a monoclinic structure of the space group $C$2/$m$ with lattice 
parameters of $a$ = 10.271 \AA, $b$ = 5.911 \AA, $c$ = 7.711 \AA, $\beta$ = 
116.42$^{\circ}$~\cite{14}. 
This structure consists of 
Cu$_3$O$_6$(OH)$_2$ layers, made up of edge-shared CuO$_4$(OH)$_2$ octahedra 
and separated by VO$_4$ tetrahedra and Ba$^{2+}$ ions (Fig. \ref{F1} (a)). 
Cu$^{2+}$ ions form a nearly perfect kagome lattice, though there 
are two crystallographic sites for them, as in volborthite. 
The distortion of a Cu triangle from the regular one is negligible ($\sim$ 0.2\%) 
with the distances between two Cu atoms being 2.962 \AA \ (Cu1-Cu2) and 
2.956 \AA \ (Cu2-Cu2)~\cite{14}. 
Thus, the spatial anisotropy in $J$ may be much smaller than that in volborthite. 
Moreover, vesignieite is expected to be free of antisite disorder,
the same as volborthite, 
because it contains no ions chemically similar to Cu$^{2+}$. 
From these structural and chemical features, we expect that vesignieite can be an ideal 
model system for the spin-1/2 KAFM to be compared with 
herbertsmithite and volborthite.

A polycrystalline sample of BaCu$_3$V$_2$O$_8$(OH)$_2$ was prepared by the hydrothermal method. 
0.1 g of the mixture of Cu(OH)$_2$ and V$_2$O$_5$ in 3:1 
molar ratio and 0.3 g of Ba(CH$_3$COO)$_2$ were put in a Teflon beaker 
placed in a stainless-steel vessel. 
The vessel was filled up to 60 volume \% with H$_2$O, sealed and 
heated at 180 $^{\circ}$C for 24 h.
Sample characterization was performed by powder x-ray diffraction (XRD) analysis 
using Cu-K$\alpha$ radiation. 
Magnetic and thermodynamic properties were measured in MPMS and PPMS (Quantum Design).
All the peaks observed in the powder XRD pattern were 
indexed to reflections based on a monoclinic structure with the lattice constants
$a$ = 10.273 \AA, $b$ = 5.907 \AA, $c$ = 7.721 \AA, $\beta$ = 116.29$^{\circ}$ 
(Fig. \ref{F1} (b)), 
confirming that our sample is single-phase vesignieite~\cite{12, 14}. 
The peaks are considerably broad, indicating a small particle size on the order 
of a few nm estimated using the Scherrer equation.

\begin{figure}
\includegraphics[width=7cm]{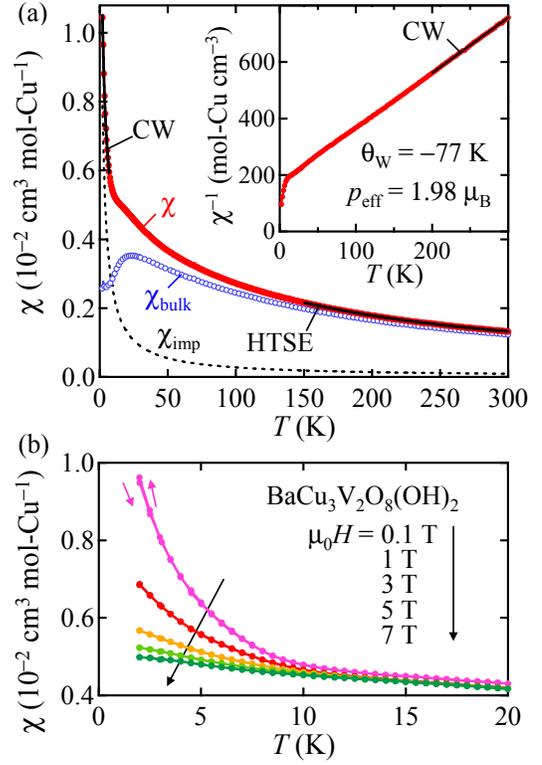}
\caption{(a) Temperature dependence of magnetic susceptibility $\chi$ of 
vesignieite BaCu$_3$V$_2$O$_8$(OH)$_2$ polycrystalline sample measured on 
heating at a magnetic field of 0.1 T. 
The filled circles, open circles, and broken line represent $\chi$, $\chi_{\textrm{bulk}}$, and
$\chi_{\textrm{imp}}$, respectively. 
The solid curve on the $\chi$ data between 2 and 10 K represents a Curie-Weiss (CW) fit, 
and that between 150 and 300 K represents a fit to the kagome lattice model obtained by
high-temperature series expansion (HTSE)~\cite{15}, 
which yields $J$/$k_{\textrm{B}}$ = 53 K and $g$ = 2.16. 
The inset shows $\chi^{-1}$, 
where the solid line between 200 and 300 K represents a CW fit, 
which gives $\theta_{\textrm{W}}$ = $-$77 K and $p_{\textrm{eff}}$ = 
1.98 $\mu_{\textrm{B}}$/Cu. 
(b) Temperature dependence of field-cooled and zero-field-cooled 
$\chi$ measured in various magnetic fields up to 5 T. 
}
\label{F2}
\end{figure}

The temperature dependences of magnetic susceptibility $\chi$ and inverse 
susceptibility $\chi^{-1}$ of 
vesignieite are shown in Fig. \ref{F2} (a). 
$\chi^{-1}$ exhibits a linear temperature dependence above 150 K, 
interpreted as Curie-Weiss magnetism. 
A Curie-Weiss fit to the data between 200 and 300 K yields a moderately large negative 
$\theta_{\textrm{W}}$ = $-$77 K and an effective moment 
$p_{\textrm{eff}}$ = 1.98 $\mu_{\textrm{B}}$/Cu, 
which is slightly larger than the spin-only value expected for $S$ = 1/2. 
The exchange coupling $J$ and Lande $g$-factor $g$ are estimated to be 
$J$/$k_{\textrm{B}}$ = 53 K and $g$ = 2.16, 
by fitting the data between 150 and 300 K to the calculation for the spin-1/2 KAFM
using the high-temperature-series expansion~\cite{15}. 
The $g$ of vesignieite is nearly equal to those of herbertsmithite 
($g$ = 2.33)~\cite{16} and volborthite ($g$ = 2.205)~\cite{3}, which are slightly larger than 2, 
reflecting a negative spin-orbit coupling constant for Cu$^{2+}$ ions. 
$J$ smaller than those of herbertsmithite ($J$/$k_{\textrm{B}}$ = 
170 K)~\cite{16} and volborthite ($J$/$k_{\textrm{B}}$ = 84 K)~\cite{3} may be
ascribed to the corresponding variation in Cu-O-Cu bond angle over the series: 
the smaller the bond angle, the larger the antiferromagnetic coupling.

\begin{figure}
\includegraphics[width=7cm]{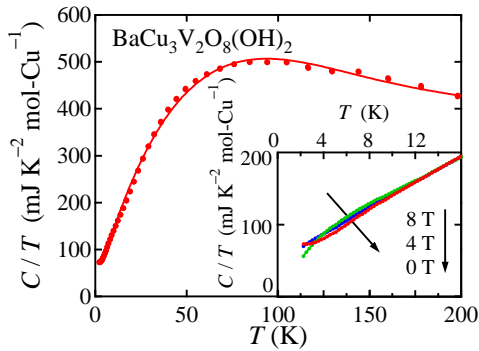}
\caption{Temperature dependence of heat capacity divided by temperature 
$C$/$T$ of vesignieite BaCu$_3$V$_2$O$_8$(OH)$_2$ polycrystalline sample 
measured at zero field. 
The inset expands the low-temperature part taken in various fields up to 8 T.
}
\label{F3}
\end{figure}

We have observed neither a long-range order nor a spin glass transition down to 2 K 
in spite of the moderately strong antiferromagnetic interaction. 
The temperature dependence of $\chi$ under various fields
shows no kink-like anomaly or thermal hysteresis between field cooling 
and zero-field cooling, as shown in Fig. \ref{F2} (b). 
The absence of long-range order is also supported by heat capacity $C$ measurements
showing no peak-like anomaly indicative of a phase transition (Fig. \ref{F3}). 
Thus, the geometrical frustration of the kagome lattice effectively suppresses 
long-range order down to low temperatures much less than $J$/$k_{\textrm{B}}$.

At low temperatures below $J$/$k_{\textrm{B}}$ = 53 K, 
$\chi$ exhibits a slightly complicated temperature dependence.
With decreasing temperature, a small hump appears around 20 K, followed
by a sharp rise below 10 K (Fig. \ref{F2} (a)). 
The latter must be due to the presence of nearly free impurity spins.
Thus, $\chi$ should contain two independent contributions:
$\chi_{\textrm{imp}}$ and $\chi_{\textrm{bulk}}$ from impurity and bulk spins, respectively.
The former may be given as 
$\chi_{\textrm{imp}}$ = $C_{\textrm{imp}}$/($T$ $-$ $\theta_{\textrm{imp}}$), 
and should increase enormously as $T$ $\rightarrow$ 0.
In contrast, the $T$ dependence of $\chi_{\textrm{bulk}}$ may be weak, 
compared with $\chi_{\textrm{imp}}$, at low temperatures.
Thus, we fit the $\chi$ data between 2 and 7 K, assuming a $T$-independent 
term $\chi_0$ instead of $\chi_{\textrm{bulk}}$ (Fig. \ref{F2} (a)), 
we obtain $C_{\textrm{imp}}$ = 
0.028 cm$^3$ K$^{-1}$ mol-Cu$^{-1}$, $\theta_{\textrm{imp}}$ = $-$1.7 K, 
and $\chi_0$ = 2.8 $\times$ 10$^{-3}$ cm$^3$ mol-Cu$^{-1}$.
$C_{\textrm{imp}}$ indicates that 7\% of spins behave as impurity spins. 
The presence of impurity spins is also evidenced in the large field dependence of 
$\chi$ observed below 10 K (Fig. \ref{F2} (b)), as well as in the small 
but finite field dependence of $C$/$T$ (the inset of Fig. \ref{F3}). 
The amount of impurity spins in vesignieite is larger than 0.07\% for volborthite but smaller
than 11\% for herbertsmithite~\cite{11, 17}, which were estimated by the
same analysis on $\chi$. 
The origin of impurity spins in vesignieite is not clear but 
may be attributed to unidentified impurity phases 
or alien spins located near the surface of small particles or crystalline defects. 

By subtracting $\chi_{\textrm{imp}}$ from $\chi$, we obtain 
the intrinsic magnetic susceptibility $\chi_{\textrm{bulk}}$, as shown in Fig. \ref{F2} (a). 
$\chi_{\textrm{bulk}}$ clearly exhibits a broad peak at 22 K $\sim$ 0.4$J$/$k_{\textrm{B}}$,
and a substantially large value of 
2.6 $\times$ 10$^{-3}$ cm$^3$ mol-Cu$^{-1}$ at $T$ = 2 K. 
Note that this broad peak has already appeared as a hump or a shoulder 
in $\chi$ and has been observed by subtracting the extrinsic 
contribution of free spins. 
The broad peak indicates that short-range order develops remarkably below 22 K, and that the
large finite value at the lowest temperature implies that the ground 
state is a gapless spin liquid, 
or, at least, that the gap is much smaller than $J$/30. 

\begin{figure}
\includegraphics[width=7cm]{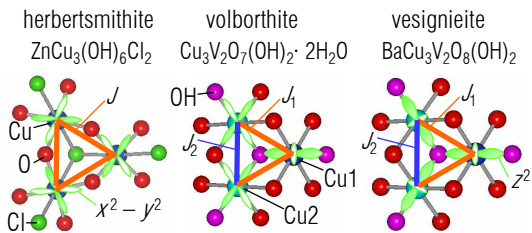}
\caption{Schematic representations of the arrangement of Cu$^{2+}$ orbitals 
in the kagome plane for 
three kagome compounds.
One of the two $e_{\textrm{g}}$ orbitals, $d_{x^2 - y^2}$ or $d_{z^2}$, carrying spin-1/2
is depicted on each Cu atom.
}
\label{F4}
\end{figure}

Let us discuss magnetic interactions in kagome lattices of the three Cu compounds
in terms of orbital arrangements.
It is common in all the three compounds that the degeneracy of the 
Cu 3$d$ $e_{\textrm{g}}$ orbitals is lifted at room temperature owing to 
the Jahn-Teller distortion. 
However, the orbital arrangements are completely different from each other among 
the three compounds, as shown in Fig. \ref{F4}, which
can be deduced reasonably from the shape of octahedra around Cu$^{2+}$ ions. 
The $z^2$ orbital is occupied by an unpaired electron
when two opposite Cu-O bonds are shorter than the other four bonds, while the 
$x^2$ $-$ $y^2$ orbital is occupied when the two Cu-O bonds are longer than the others.
In herbertsmithite, all the Cu$^{2+}$ ions have their 
unpaired electrons in the $x^2$ $-$ $y^2$ orbital,
which are placed symmetrically around the three-fold axis,
resulting in a structurally perfect kagome lattice~\cite{17}. 
In volborthite, on the other hand, the $x^2$ $-$ $y^2$ orbital is responsible for 
the Cu2 site, while the $z^2$ orbital for the Cu1 site~\cite{17}, 
which causes a disparity between $J_1$ (Cu1-Cu2) and $J_2$ (Cu2-Cu2). 
In contrast, a single orbital must be selected in vesignieite, as in the case of 
herbertsmithite, but it must be the $z^2$ orbital instead of the $x^2$ $-$ $y^2$ orbital orbital,
giving rise to an essentially undistorted kagome network (Fig. \ref{F4}). 
Nevertheless, the actual crystal symmetry is not trigonal but monoclinic, 
which comes from a 
lateral shift in the stacking of the kagome layers through intervening layers containing 
Ba$^{2+}$ ions and VO$_4$ tetrahedra.
Since the distortion of Cu triangles in vesignieite 
is negligible, only 0.2\%, one can consider vesignieite as a nearly ideal kagome system,
which is better than volborthite from a structural point of view, 
and cleaner than herbertsmithite in terms of impurity spins. 

\begin{figure}[tb]
\includegraphics[width=7cm]{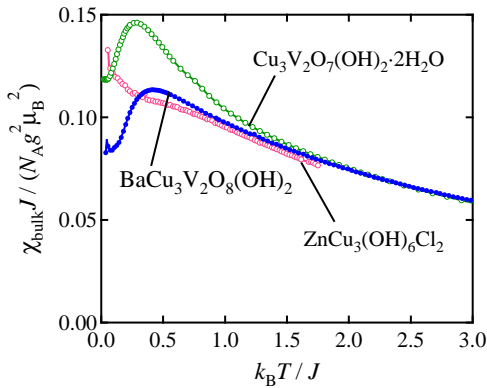}
\caption{Temperature dependence of normalized 
$\chi_{\textrm{bulk}}$ for vesignieite BaCu$_3$V$_2$O$_8$(OH)$_2$, 
volborthite Cu$_3$V$_2$O$_7$(OH)$_2$$\cdot$2H$_2$O and 
herbertsmithite ZnCu$_3$(OH)$_6$Cl$_2$. 
The $J$/$k_{\textrm{B}}$ and $g$ used are 53 K and 2.16 
for vesignieite, 84 K and 2.21 for volborthite~\cite{3}, 
and 199 K and 2.23 for herbertsmithite~\cite{17}, respectively.
}
\label{F5}
\end{figure}

Finally, we compare the $\chi_{\textrm{bulk}}$ for 
the three kagome compounds, 
in order to deduce the general feature of the KAFM.
The temperature dependence of $\chi_{\textrm{bulk}}$ normalized by both $J$ and $g$ is 
shown in Fig. \ref{F5}. 
At a high temperature, all the curves tend to merge, 
as expected from the mean-field picture. 
However, they separate from each other on 
cooling below $T$ $\sim$ 1.5$J$/$k_{\textrm{B}}$ and show broad peaks 
at $T_{\textrm{peak}}$ = 0.4$J$/$k_{\textrm{B}}$ for vesignieite 
and 0.25$J$/$k_{\textrm{B}}$ for volborthite, 
indicative of the formation of short range order. 
Although the presence of such a broad peak is not discernible for herbertsmithite,
this may be because of the experimental ambiguity in estimating
a large $\chi_{\textrm{imp}}$ in $\chi$. 
In fact, in a recent NMR study of herbertsmithite, it was found that local 
susceptibility decided from shift measurements, which may be unaffected by impurity spins
or defects in the kagome plane, exhibits a similar broad maximum at around 0.4$J$~\cite{19}.
Therefore, $T_{\textrm{peak}}$, as well as the magnitude of $\chi_{\textrm{bulk}}$, are in 
good agreement between vesignieite and herbertsmithite, suggesting a general feature
of the KAFM.
In contrast, the normalized $\chi_{\textrm{bulk}}$ of volborthite seems considerably 
different from the others: $T_{\textrm{peak}}$ is lower, 
and the magnitude of $\chi_{\textrm{bulk}}$ at $T_{\textrm{peak}}$ is significantly larger.
This is likely due to the spatial anisotropy in $J$. 
Recent theoretical calculations of $\chi$ on a spin-1/2 spatially 
anisotropic Heisenberg model on the kagome lattice successfully reproduced these tendencies
with increasing anisotropy~\cite{18}.

Common to all the three compounds, $\chi_{\textrm{bulk}}$ seems to approach a large finite value at
$T$ = 0, indicative of the absence of a spin gap having a realistic magnitude.
Therefore, the intrinsic ground state of the spin-1/2 KAFM must be a 
gapless spin liquid state or a state with a gap much smaller 
than theoretically predicted~\cite{1}.
This unusual ground state seems to be robust to the spatial anisotropy of the kagome lattice, 
as observed in volborthite.
It is important to determine whether the anomalous magnetization steps 
that have recently been observed in the
high-quality sample of volborthite~\cite{11} are general features of the KAFM 
and are also observed in vesignieite. 
To answer this question, a higher-quality sample with fewer impurity spins is required,
and an effort to improve sample quality is in progress.

In conclusion, we have demonstrated that vesignieite BaCu$_3$V$_2$O$_8$(OH)$_2$ is a spin-1/2 
antiferromagnet on a nearly ideal kagome lattice. 
The kagome lattice in vesignieite is more spatially isotropic than that in volborthite, 
and contain fewer impurity spins than that in herbertsmithite. 
The $\chi_{\textrm{bulk}}$ of vesignieite shows a broad peak 
at $\sim$ 0.4$J$/$k_{\textrm{B}}$ due to short-range ordering and a large finite value as
$T$ $\rightarrow$ 0, without a long-range order, a spin glass transition and a spin gap down to
$J$/30. These features are common to other kagome compounds, 
suggesting that the ground state of the spin-1/2 KAFM is a very small 
gapped or gapless spin liquid.

This work was partly supported by a Grant-in-Aid for Scientific Research on 
Priority Areas ``Novel States of Matter Induced by Frustration'' (No. 19052003).

\end{document}